\documentclass{article}
\usepackage[T2A]{fontenc}
\usepackage[russian]{babel}
\usepackage{graphicx}
\usepackage{amsbsy}
\usepackage{amssymb}
\begin{document}
\makeatletter \Large \baselineskip 6mm \centerline{\bf Dynamic Characteristics of the Low-Temperature Decomposition}
\centerline{\bf of the C$_{20}$ Fullerene}

\vskip 4mm

\centerline{K.P. Katin$^{1)}$, A.I. Podlivaev$^{2)}$}

\vskip 4mm

\centerline{\it National Nuclear Research University MEPhI}
\centerline{\it Kashirskoe sh. 31, Moscow, 115409 Russia}

\vskip 5mm

$^{1)}$e-mail: KPKatin@gmail.com \vskip 1mm $^{2)}$e-mail: AIPodlivayev@mephi.ru

\vskip 5mm

\centerline{\bf Abstract}

A novel algorithm has been proposed for simulating thermal decomposition of atomic clusters at
such low temperatures that the corresponding lifetimes are macroscopic and, hence, standard molecular
dynamics algorithms are inapplicable. The proposed algorithm is based on a combination of the molecular
dynamics and Monte Carlo techniques. It is used to calculate the temperature dependence of the lifetime of
the thermalized C$_{20}$ fullerene until it decomposes at $T$ = 1300$-$4000 K. The frequency factor and activation
energy of the decomposition are determined. It is demonstrate that the temperature dependences of the lifetimes of the heat-isolated and thermalized fullerenes differ significantly.

\vskip 5mm


\vskip 8mm

Currently, numerical simulations of the evolution of small atomic clusters are done using the molecular
dynamics (MD) or Monte Carlo (MC) techniques. When the molecular dynamics method is used to simulate atomic clusters, the limited performance of modern computers limits the time period during which the ''life'' of the cluster can be followed. Even for relatively small systems composed of fewer than 100 atoms, this interval is limited to several tens of picoseconds for calculations using ab initio methods (see, for example,~\cite{bib:1,bib:2}), and to several microseconds for simplified techniques using classical interatomic interaction poten\-tials~\cite{bib:3,bib:4} or tight-binding potentials~\cite{bib:5,bib:6,bib:7}. This results, in particularly, in the inability to investigate the stability of clusters at relatively low (corresponding to the real experimental conditions) temperatures $T \sim $ 1000 K, because the cluster lifetime τ increases exponentially with lowering temperature, and, accordingly, the duration required to perform calculations (''computational experiment'') expo\-nentially increases.

In some cases, the problem of exponentially decreasing rate of evolution processes in a cluster with lowering temperature is efficiently solved within various modifica\-tions of the kinetic Monte Carlo approach~\cite{bib:8,bib:9} involving the transition state theory (TST)~\cite{bib:10}. The Monte Carlo + transition state theory approach offers a way to substantially increase the performance of the Monte Carlo technique as compared to the molecular dynamics method, which eliminates the problem of time scaling of dynamic processes, and makes it possible to observe the evolution of the cluster over macroscopic time spans~\cite{bib:11}. This approach, however, is only applicable to a rather narrow class of problems, in which the many-particle potential of the system in the generalized atomic coordinate space of the cluster has a set of minima (with the known residence times of the system), and the evolution of the system is determined by transitions between these minima. This class includes, in particular, the problems of island formation by atoms adsorbed onto the surface of a crystalline adsorbent, the diffusion and coagulation of interacting defects in solids, and so on; however, the problem of decomposition of an isolated cluster doesn’t belong to this class. For example, moving along the reaction coordinate from a perfect (defect-free) C$_{20}$ fullerene towards the products of its decomposition, only four or five metastable configurations can be detected~\cite{bib:12}, which is insufficient for the efficient use of the Monte Carlo + transition state theory approach.

The combined Monte Carlo + molecular dynamics algorithm has been proposed for simulations of atomic clusters on prolonged time scales~\cite{bib:13}. The evolution of the system is described by the stochastic motion of atoms according to the Metropolis Monte Carlo algorithm~\cite{bib:14}. Each step of this algorithm corresponds to a certain time interval. The correspondence between one step of the Metropolis algorithm and the real (''physical'') time interval is determined from matching the atomic diffusion mobilities computed using the molecular dynamics method and the Monte Carlo approach. However, the performan\-ce of the algorithm~\cite{bib:13} offers only a fourfold to sixfold advantage with respect to the molecular dynamics method, which obviously is not sufficient. Apart from the relatively low performance, another shortcoming of the combined algorithm~\cite{bib:13} is that it is applicable only to some dynamical problems, since the Metropolis algorithm (the most commonly used among the Monte Carlo methods) has been developed for calculations of the characteristics of stationary statistical equilibrium systems. The fundamental justification of this algorithm in~\cite{bib:14} is based on the detailed equilibrium principle, which, strictly speaking, is not valid for the irreversible process of the cluster decomposition (when the statistics is accumulated for these problems, only the probability flow for the transition of the system from the initial cluster to the products of its decomposition is taken into consideration, while the reverse process of ''regeneration'' of the cluster is ignored, which violates the detailed equilibrium principle). The algorithm proposed in the present work is based on the idea of the sequential application of the molecular dynamics and Monte Carlo methods~\cite{bib:13}, but it is modified so as to remediate the principal shortcomings of the technique~\cite{bib:13}.

In our approach, the region in the configuration space, where the potential energy of the cluster deviates weakly from its average value, and the detailed equilibrium condition is satisfied well enough, is simulated using the Monte Carlo technique (the Metropolis algorithm~\cite{bib:14}). The regions with a higher potential energy (exceeding a certain threshold value $U$) are termed fluctuation regions in the following. The dynamics of the system in these regions is simulated using the molecular dynamics method. The magnitude of the threshold potential energy U is an important parameter in the proposed scheme. The correct choice of the value of the threshold potential energy makes it possible to eliminate the systematic error of the Metropolis algorithm caused by the violation of the detailed equilibrium principle. In our algorithm, the Monte Carlo step does not correspond to a fixed time interval, as in~\cite{bib:13}. The time scale matching is done basing on the average time interval between two sequential fluctuations, which is determined in advance using the molecular dynamics method. The proposed algorithm is multistep. During the simulation of the system with a lower fluctuation threshold of the potential energy $U$, the average time interval between fluctuations with a threshold $U_1$ > $U$ is determined, and, when the accumulated statistics is sufficient, the simulation is repeated with a new, larger threshold value $U_1$. The increase of the threshold value $U$ is repeated up to the decomposition of the cluster. Owing to its multistep structure, our algorithm outperforms the molecular dynamics method by many orders of magnitude, and is comparable to the Monte Carlo + transition state theory approach, while at the same time being free of the limitations that the transition state theory imposes on the potential energy landscape of the cluster.

In our simulations of the decomposition of the C$_{20}$ fullerene, we used the tight-binding potential~\cite{bib:6} which performs adequately in the description of the dynamics of thermal fragmentation of the C$_{60}$ fullerene~\cite{bib:15}, the Stone-Wales transformation activation energy in the C60 fullerene~\cite{bib:16}, as well as the various covalent bond types in cluster molecules~\cite{bib:17}, linear chains~\cite{bib:18,bib:19}, and two-dimensional complexes~\cite{bib:20} of the C$_{20}$ fullerenes. We investigated the characteris\-tics of the decomposition of the thermalized C$_{20}$ fullerene and determined its lifetime $\tau$ in the 1300-4000 K temperature range. The dependence of the logarithmic lifetime on the inverse temperature of the cluster is shown in the figure. It can be seen from this figure that the use of the new algorithm makes it possible to simulate the process of cluster decomposition down to the temperature of $T$ = 1300 K, when its lifetime reaches macroscopic values on the order of one second. Such a broad temperature range allows us to determine the activation energy $E_a$ and the frequency factor $A_0$ of the decomposition of the thermalized C$_{20}$ fullerene in the Arrhenius equation for the cluster lifetime,
\begin{equation}
1/\tau(T)=A_0 exp^{-E_a/kT},
\end{equation}
with the precision previously inaccessible for numerical methods. These quantities are found to be equal to $E_a = 4.98 \pm 0.16$ eV and $A_0 = (2.9 \pm 0.3)\cdot 10^{17} s^{-1}$, correspondingly. We note that the activation energy coincides within the error margin with the minimum potential barrier height (5.0 eV), which was calculated from the analysis of the potential energy of the cluster as a function of the coordinates of all constituent atoms~\cite{bib:12}. Further, the $A_0$ magnitude is dramatically different from the frequency factor of the decomposition of the thermalized C$_{60}$ fullerene, $A = (1.1 \pm 0.1)\cdot 10^{21} s^{-1}$, as determined using the molecular dynamics methods~\cite{bib:21}. It should also be noted that the activation energy in~\cite{bib:21}, i.e., $E_a = 6.4 \pm 0.4$ eV, turns out larger than the minimum potential barrier height. The substantial discrepancies between the frequency factors and between the activation energies could result from both the insufficiently broad temperature region (2400-4000 K) accessible for conventional the molecular dynamics calcula\-tions~\cite{bib:21}, as well as the differences between the functional form of the temperature dependen\-ces of the lifetimes of the thermalized and isolated fullerenes. Indeed, the shape of the dependence of the decomposition lifetime for an isolated cluster on the average kinetic energy of its atoms differs from the Arrhenius form~\cite{bib:22}, and, therefore, the pre-exponential factor may not coincide with the frequency factor for the decomposition of a thermalized cluster put into contact with the surrounding medium.

We’d like to stress the fact that our algorithm makes it possible not only to determine the lifetime of the cluster until its decomposition, but also, to study the scenario of the decomposition and its products. The analysis of defect configurations generated throughout the evolution process of the therma\-lized C$_{20}$ fullerene shows that they correspond precisely to the configurations observed in an isolated fullerene~\cite{bib:20} (these defect configurations possess, aside from pentagons
that constitute a perfect C$_{20}$ fullerene, large adjacent ''windows'' on the surface).

The full correspondence of the decomposition scenarios for isolated and thermalized fullerenes, the agreement of the activation energy of the decomposition with the height of the energy barrier, and the nearly perfect linear dependence of the logarithm of the lifetime on the inverse temperature (see the figure) demonstrate that our approach is applicable to simulations of long-lived metastable atomic systems. The discrepancy between temperature dependences of the lifetimes for isolated and therma\-lized clusters should necessarily be taken into account during the analysis of the experimental data from, on the one hand, the rapid photofrag\-mentation of clusters in a rare gas medium, and the slow decomposition in a dense medium, on the other.

We would like to thank L.A. Openov for useful discussions. This study was supported by the Russian Foundation for Basic Research (project no. 09-02-00701-a) and performed within the Analytical Department Target Program ''Development of the Scientific Potential of the Higher School (2009-2010)'' (project no. 2.1.1/468).

\renewcommand{\refname}{\begin{center}{\Large\rm\bf REFERENCES}\end{center}}

\newpage
\includegraphics[width=15cm,height=15cm]{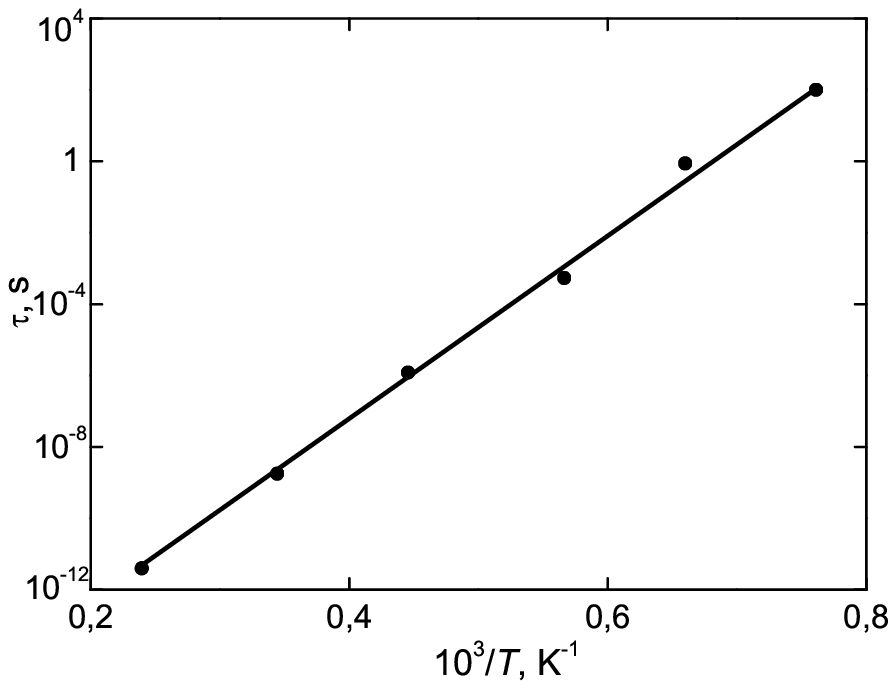}
\vskip 20mm
{\bf Fig.1.} Dependence of the lifetime $\tau$ of the thermalized C$_{20}$ fullerene on the inverse temperature. Points represent the results of the averaging over four independent stages of the simulation at each of the following temperatures: $T$ = 4000, 2824, 2182, 1777, 1500, and 1300 K. Vertical bars indicate the dispersion at the corresponding temperature. The solid line corresponds to the linear least-squares interpolation over the six average lifetimes.

\end{document}